\documentclass[altaffilletter,aps,nofootinbib,twocolumn,prd,eqsecnum,preprintnumbers,superscriptaddress]{revtex4-1}
\pdfoutput=1
\usepackage[caption=false]{subfig}
\usepackage{graphicx}
\usepackage{amsmath}
\usepackage{amsfonts}
\usepackage{amssymb}
\usepackage{color}
\usepackage{bm}
\usepackage{mathrsfs}
\usepackage{epstopdf}
\usepackage{url}
\usepackage{footnote}
\usepackage{textcomp}
\usepackage{dsfont}
\usepackage{ulem}
\usepackage{hyperref}
\usepackage{enumerate}   
\usepackage{appendix}
\usepackage{textcomp}
\usepackage{tipa}
\begin{document}

\title{Modified Teleparallel Gravity induced by quantum fluctuations}

\author{Che-Yu Chen}
\email{b97202056@gmail.com}
\affiliation{Institute of Physics, Academia Sinica, Taipei 11529, Taiwan}

\author{Yu-Hsien Kung}
\email{r06222010@g.ntu.edu.tw}
\affiliation{Physics Division, National Center for Theoretical Sciences, Taipei 10617, Taiwan}
\affiliation{Department of Physics and Center for Theoretical Sciences, National Taiwan University, Taipei 10617, Taiwan}

\begin{abstract}
In the semi-classical regime, quantum fluctuations embedded in a Riemannian spacetime can be effectively recast as classical back reactions and manifest themselves in the form of non-minimal couplings between matter and curvature. In this work, we exhibit that this semi-classical description can also be applied within the teleparallel formulation. In the teleparallel formulation, quantum fluctuations generically lead to non-minimal torsion-matter couplings. Due to the equivalence between the (classical) Einstein gravity in the Riemannian description and that in teleparallel description, some effective models which were constructed using Riemannian description can be reproduced completely using the teleparallel description. Besides, when the effective quantum correction term is proportional to the torsion scalar $T$, we obtain a subclass of novel $f(T,B,\mathcal{T})$ gravity, where $B$ is a boundary term, and $\mathcal{T}$ is the trace of the energy-momentum tensor. Next, we investigate the cosmological properties in this $f(T,B,\mathcal{T})$ theory by assuming that the matter  Lagrangian is solely constructed by a dynamical scalar field. We exhibit some interesting cosmological solutions, such as those with decelerating expansion followed by a late-time accelerating phase. In addition, the non-minimal torsion-matter couplings induced by quantum corrections naturally lead to energy transfers between gravity and cosmological fluids in the universe.

\end{abstract}

\maketitle

\section{Introduction}

Although Einstein's General Relativity (GR) is very successful in describing our universe, it still suffers from several essential puzzles. From the theoretical point of view, it is still not clear how to consistently embed quantum effects into the framework of GR. In fact, the consistent formulation of a fundamental quantum theory of gravity is still lacking. From the observational perspectives, on the other hand, the observations of recent accelerating expansion of the universe, as well as the mysterious existence of some ``unseeable" matter fields clustering around galaxies and galaxy clusters, challenge our current understanding of cosmology. The simplest explanation of these exotic phenomena may be the inclusion of some non-standard matter fields, such as dark energy and dark matter, respectively. However, some problems still remain. For example, the questions that where these dark sectors come from, why the amount of dark energy is so tiny, and why the relative amount of dark energy and dark matter is so fine-tuned, are still open issues.    

In order to address the aforementioned theoretical and observational issues, one phenomenological approach is to consider modified theories of GR \cite{Nojiri:2017ncd}. In particular, these modified theories of gravity are expected to not only retain the success that has been achieved by GR, but also shed some light on ameliorating the fundamental puzzles mentioned above. One important ingredient within this framework is how to motivate the idea of modified theories of gravity from quantum origins. Such a connection can indeed be realized, at least, within the semi-classical regimes. In fact, in quantum mechanical descriptions of gravity, one can naively replace all the classical quantities in GR by some quantum operators. In this regard, the classical Einstein equation can then be replaced by an operator equation:
\begin{equation}
\hat{G}_{\mu\nu}=\frac{8\pi G}{c^4}\hat{\mathcal{T}}_{\mu\nu}\,,\label{quantumeheq}
\end{equation} 
where $G_{\mu\nu}$ and $\mathcal{T}_{\mu\nu}$ are the Einstein tensor and the energy-momentum tensor, respectively. $G$ and $c$ are the gravitational constant and the speed of light. The hat denotes the operator of the field. This operator equation is in general very difficult to solve, but it could be dealt with in the semi-classical regimes based on some assumptions about the expectation values of metric operators. 

In Refs.~\cite{Dzhunushaliev:2013nea,Dzhunushaliev:2015mva}, the authors proposed an idea to solve Eq.~\eqref{quantumeheq} within the semi-classical approximations. It was assumed that, in the semi-classical approximations, the quantum metric operator $\hat{g}_{\mu\nu}$ can be decomposed into a classical part and a quantum fluctuating part. They further assumed that the expectation value of the quantum fluctuating part of the metric is non-zero and is given by a rank-two tensor built by some classical entities. Then, considering the expansion only up to the first-order of the quantum fluctuating metric, some effective models can be constructed \cite{Dzhunushaliev:2019jpg}. Interestingly, the resulting effective theories are commonly featured by the existence of non-minimal curvature-matter couplings. These non-minimal couplings are purely quantum in nature, and one direct consequence of these couplings is the non-conservation of the energy-momentum tensor, i.e., $\nabla_\mu \mathcal{T}^{\mu\nu}\ne0$. Therefore, these non-minimal curvature-matter couplings can be naturally interpreted as some irreversible particle creation processes, or more explicitly, some energy transfer between the gravitational sector and the matte sector. In fact, such particle creation processes are quite common when considering quantum field theory in curved spacetimes, in particular in the paradigm of cosmological vacuum \cite{Parker:2009uva}. The cosmological solutions \cite{Yang:2015jla,Liu:2016qfx} as well as the black hole solutions \cite{Yang:2020lxv} of this effective approach have been investigated.

The idea of introducing non-minimal matter-geometry couplings is actually not new in the community of modified theories of gravity. In the literature, several modified theories of gravity with non-minimal matter couplings have been proposed in order to explain the late-time accelerating expansion of the universe. For example, one may couple the Ricci scalar $\mathring{R}$ non-minimally to the matter Lagrangian $\mathcal{L}_m$, to the energy-momentum tensor, or to a dynamical scalar field $\phi$, giving rise to, for example, the $f(\mathring{R},\mathcal{L}_m)$ gravity \cite{Bertolami:2007gv,Harko:2008qz,Harko:2010mv,Harko:2020ibn}, $f(\mathring{R},\mathcal{T})$ gravity \cite{Harko:2011kv}, $f(\mathring{R},\mathcal{T},\mathring{R}_{\mu\nu}\mathcal{T}^{\mu\nu})$ gravity \cite{Haghani:2013oma,Odintsov:2013iba}, and the energy-momentum squared gravity \cite{Katirci:2013okf,Roshan:2016mbt}, where $\mathcal{T}$ is the trace of the energy-momentum tensor. It should also be pointed out that by choosing a proper quantum fluctuating metric, the semi-classical approach proposed in Refs.~\cite{Dzhunushaliev:2013nea,Dzhunushaliev:2015mva} can lead to effective theories that belong to the $f(\mathring{R},\mathcal{T})$ gravity or theories that simultaneously contain $\mathcal{T}_{\mu\nu}\mathcal{T}^{\mu\nu}$ and $\mathring{R}_{\mu\nu}\mathcal{T}^{\mu\nu}$ couplings \cite{Liu:2016qfx}.

Besides the Riemannian formulation in which the fundamental field is the metric tensor and the formulation is based on the standard Riemannian spacetime, in which gravity is described by spacetime curvatures, one can also construct gravitational theories using the so-called teleparallel formulation. In the teleparallel formulation, the fundamental fields are the tetrad field $e_\mu^a$ and the spin connection ${\omega^a}_{b \mu}$, where the former encodes the spacetime degrees of freedom and the latter represents the Lorentz degrees of freedom. The spacetime is assumed to have zero curvature, and the gravitational effects are completely described by the torsion, which is constructed by the tetrad and spin connection \cite{deAndrade:1997gka,deAndrade:2000kr,Obukhov:2002tm}. Using the torsion scalar $T$ to construct the gravitational Lagrangian, one can obtain an equivalent description of GR, i.e., the teleparallel equivalent of general relativity (TEGR) \cite{Maluf:2013gaa}, in the sense that they have the same equations of motion, but the geometrical interpretations are different. For the most updated review of teleparallel gravity and its cosmological applications, we refer the readers to Ref.~\cite{Bahamonde:2021gfp}.

Similar to how one can modify GR within the Riemannian formulation, one can also go beyond TEGR within the teleparallel formulation. For example, even though GR and TEGR are equivalent with the actions of the former and the latter defined by the Ricci scalar $\mathring{R}$ and the torsion scalar $T$, respectively, their generalizations, for example, the $f(\mathring{R})$ and $f(T)$ gravity \cite{Ferraro:2006jd,Cai:2015emx,Krssak:2018ywd} are completely different theories. The $f(\mathring{R})$ gravity can only be recovered when considering a particular subclass of the general $f(T,B)$ gravity \cite{Bahamonde:2015zma}, where $B$ is a boundary term that satisfies $\mathring{R}=B-T$. Similar to the theories with non-minimal curvature-matter couplings in the Riemannian formulation, one can also construct modified theories containing non-minimal torsion-matter couplings in the teleparallel formulation. For instance, direct non-minimal couplings between the torsion and a dynamical scalar filed were proposed in Ref.~\cite{Geng:2011aj} and some interesting dark energy phenomenologies were found. This model was further extended to include couplings with the boundary term \cite{Bahamonde:2015hza}. In addition, one can couple the torsion scalar non-minimally to the matter Lagrangian or to the trace of the energy-momentum tensor, giving rise to the $f(T,\mathcal{L}_m)$ gravity \cite{Harko:2014sja} and $f(T,\mathcal{T})$ gravity \cite{Harko:2014aja}, respectively. Very recently, the theory with general non-minimal torsion-matter-boundary couplings was proposed, i.e., the $f(T,B,\mathcal{L}_m)$ gravity \cite{Bahamonde:2017ifa}. The cosmological solution was investigated using the dynamical analysis. In addition, it has been shown that the $f(T,\mathcal{L}_m)$ as well as the $f(\mathring{R},\mathcal{L}_m)$ can be recovered as subclasses of this general theory. In the literature, other generalizations in the teleparallel framework have also been proposed, such as by formulating it in the Horndeski approach \cite{Bahamonde:2019shr}, including a Gauss-Bonnet coupling \cite{Kofinas:2014owa,Bahamonde:2016kba}, or coupling the theory with vector fields \cite{Motavalli:2018ien}. These theories, in particular, their cosmological applications have been widely investigated as well \cite{Skugoreva:2014ena,Farrugia:2016qqe,Oikonomou:2016jjh,RezaeiAkbarieh:2018ijw,Bernardo:2021izq,Bernardo:2021bsg,Abedi:2018lkr,Abedi:2015cya}.
   
The fact that the irreversible particle creation processes appear both in modified theories of gravity with non-minimal matter couplings and in quantum field theories in curved spacetimes suggests that there may be a deeper relation between them \cite{Harko:2012za}. The aim of this paper is to address the following question: Is the non-minimal geometry-matter couplings generic enough in gravitational theories with quantum corrections? Based on the results of Refs.~\cite{Dzhunushaliev:2013nea,Dzhunushaliev:2015mva}, one may conclude that non-minimal curvature-matter couplings would appear naturally when quantum effects are included in the standard Riemannian formulation. In this work, we will follow the similar idea of Refs.~\cite{Dzhunushaliev:2013nea,Dzhunushaliev:2015mva}. More explicitly, we will adopt the teleparallel formulation and assume that the tetrad field is quantum in nature. Then we will decompose the tetrad operator into a classical part and a quantum part. In the semi-classical level, we will show that the effective theories obtained using this method are featured by the non-minimal couplings between torsion and matter fields. In particular, by assuming that the expectation value of the quantum tetrad is proportional to the torsion scalar, the resulting effective theory turns out to be the subclass of a general $f(T,B,\mathcal{T})$ gravity. The effective theory contains non-minimal torsion-boudary-matter couplings, and it also contains higher-order derivatives in the equations of motion. The latter naturally appears in the teleparallel gravitational theories coupled with the boundary term $B$. We will investigate the cosmological equations of this novel effective theory and show the possibilities of having some interesting cosmological solutions.

This paper is outlined as follows. In Sec.~\ref{sec.TEGR}, we briefly review the teleparallel formulation and explain how one can obtain an equivalent description of GR within the teleparallel formulation. In Sec.~\ref{sec.Qtegr}, we extend the semi-classical quantum gravitational approach of Refs.~\cite{Dzhunushaliev:2013nea,Dzhunushaliev:2015mva} to teleparallel formulation, and construct the effective teleparallel Lagrangian in its general form with a general quantum tetrad. Some particular examples of quantum tetrads are chosen in Sec.~\ref{sec.example} and the corresponding effective Lagrangians are given. The cosmological solutions of a particular novel effective theory with non-minimal torsion-matter-boundary couplings are investigated in Sec.~\ref{sec:cosmology}. We finally conclude in Sec.~\ref{sec.con}.

\section{Teleparallel formalism of gravity}\label{sec.TEGR}
In this section, we will briefly review the formulation of teleparallel gravity and why an equivalent description of GR can be formulated within the teleparallel framework, namely, the teleparallel equivalent of general relativity (TEGR). In teleparallel formulation, one fundamental field is the tetrad $e_\mu^a$, which is an orthonormal basis that maps the local spacetime point $x^\mu$ to the point $x^a$ on the tangent bundle, or the soldered bundle \cite{Lucas:2008gs}. Using the tetrad, the spacetime metric can be expressed as
\begin{align}
    g_{\mu\nu} = \eta_{ab} e_\mu^a e_\nu^b \,,
\end{align}
where $\eta_{ab} = \text{diag}(1,-1,-1,-1)$.
Note that the Greek indices are spacetime indices, and Latin indices are tangent bundle indices. In TEGR, gravitational effects can be alternatively described by torsions, rather than curvatures. Different from the Riemann tensor in GR that is constructed  by the torsionless Levi-Civita connection, the torsion tensor is constructed by the teleparallel connection $\Gamma^\lambda_{\nu\mu}$
\begin{align}
    {T^\lambda}_{\mu\nu} &= -2 \Gamma^\lambda_{[\mu\nu]} \nonumber\\&= e_a^\lambda \partial_\mu e^a_\nu - e_a^\lambda \partial_\nu e^a_\mu + e_a^\lambda {\omega^a}_{b\mu} e^b_\nu - e_a^\lambda {\omega^a}_{b\nu} e^b_\mu
    \,,
\end{align}
where the teleparallel connection $\Gamma^\lambda_{\nu\mu}$ is defined by
\begin{align}
    \Gamma^\lambda_{\nu\mu} = e_a^\lambda \partial_\mu e^a_\nu + e_a^\lambda {\omega^a}_{c\mu} e^c_\nu
    \,.
\end{align}
Note that ${\omega^a}_{c\mu}$ is a flat spin connection which contributes to the local Lorentz degrees of freedom in TEGR. 

Based on the definitions of the teleparallel connection, one can write down the relation between the Levi-Civita connection $\mathring{\Gamma}^\lambda_{\mu\nu}$ and the teleparallel connection as follows: \cite{Aldrovandi:2013wha}
\begin{align}
\label{relation}
   \Gamma^{\lambda}_{\mu\nu} = \mathring{\Gamma}^\lambda_{\mu\nu} + {K^\lambda}_{\mu\nu} \,,
\end{align}
where
\begin{align}
\label{contorsion}
    {K^\lambda}_{\mu\nu} = \frac{1}{2} \left( {{T_\nu}^\lambda}_\mu + {{T_\mu}^\lambda}_\nu - {T^\lambda}_{\mu\nu}  \right) \,,
\end{align}
is the contorsion tensor. The relation \eqref{relation} between $\mathring{\Gamma}^\lambda_{\mu\nu}$ and $\Gamma^\lambda_{\mu\nu}$ is the fundamental reason why teleparallel formulation can describe gravity with torsion. Furthermore, one can directly verify the equivalence between GR and TEGR by studying the relation between their actions.
The  Lagrangian density of TEGR can be written as
\begin{align}
\label{Ltegr}
    \mathcal{L} = \frac{e}{2\kappa^2} T + e \mathcal{L}_m \,,
\end{align}
where $e = \text{det} ( e_\mu^a) = \sqrt{-g}$, $\kappa^2 = 8 \pi G/c^4 $, and $\mathcal{L}_m$ is the matter Lagrangian density. The torsion scalar is defined by $T \equiv K^{\mu\nu\rho} K_{\rho\nu\mu} - {K^{\mu\rho}}_\mu {K^\nu}_{\rho\nu} = T_{\rho\mu\nu} S^{\rho\mu\nu}$, where the superpotential $S^{\rho\mu\nu}$ is defined by
\begin{align}
    S^{\rho\mu\nu} &\equiv \frac{1}{2} \left( K^{\mu\nu\rho} - g^{\rho\nu} {T^{\sigma\mu}}_\sigma + g^{\rho\mu} {T^{\sigma\nu}}_\sigma \right)
    \,.
\end{align}
To show the equivalence between the Einstein-Hilbert Lagrangian and the TEGR Lagrangian \eqref{Ltegr}, we may first use Eq.~\eqref{relation} to obtain the relation between the standard Riemann curvature ${\mathring{R}^\rho}_{\,\,\,\theta\mu\nu}$ and the curvature constructed by the teleparallel connections ${R^{\rho}}_{\theta\mu\nu}$, which is supposed to be zero in the tangent bundle.

Therefore, we have
\begin{align}
\label{R+N=0}
    {R^\rho}_{\theta\mu\nu} \equiv {\mathring{R}^\rho}_{\,\,\,\theta\mu\nu} + {N^\rho}_{\theta\mu\nu} = 0 \,,
\end{align}
where
\begin{align}
     &{N^\rho}_{\theta\mu\nu} = \partial_\mu {K^\rho}_{\theta\nu} - \partial_\nu {K^\rho}_{\theta\mu} + \Gamma^{\rho}_{\sigma\mu} {K^\sigma}_{\theta\nu} - \Gamma^{\rho}_{\sigma\nu} {K^\sigma}_{\theta\mu} 
     \nonumber\\ 
     &- \Gamma^{\sigma}_{\theta\mu} {K^\rho}_{\sigma\nu} + \Gamma^{\sigma}_{\theta\nu} {K^\rho}_{\sigma\mu} + {K^\rho}_{\sigma\nu} {K^\sigma}_{\theta\mu} - {K^\rho}_{\sigma\mu} {K^\sigma}_{\theta\nu} 
\end{align}
is a tensor written in terms of the teleparallel connection only. The relation \eqref{R+N=0} implies $ - {\mathring{R}^\rho}_{\,\,\,\theta\mu\nu} = {N^\rho}_{\theta\mu\nu}$. By taking appropriate contractions, one can obtain the relation between the torsion scalar $T$ and the standard Ricci scalar $\mathring{R}$ as follows
\begin{align}
\label{RQ}
    - \mathring{R} &= N = \left(  K^{\mu\nu\rho} K_{\rho\nu\mu} - {K^{\mu\rho}}_\mu {K^\nu}_{\rho\nu}  \right) + \frac{2}{e} \partial_\mu \left( e {T^{\nu\mu}}_\nu \right)
    \nonumber \\
    &= T - B \,,
\end{align}
where $B=- 2 \partial_\mu \left( e {T^{\nu\mu}}_\nu \right)/e$ is a boundary term. The standard Ricci scalar $\mathring{R}$ is thus different from the torsion scalar by a boundary term $B$. Therefore, the relation between GR and TEGR Lagrangians can be expressed as
\begin{align}
    \frac{e}{2\kappa^2}T = \mathcal{L}_{\textrm{EH}} + \frac{e}{2\kappa^2} B \,,
\end{align}
where
\begin{align}
   \mathcal{L}_{\textrm{EH}} = - \frac{\sqrt{-g}}{2\kappa^2}  \mathring{R} 
\end{align}
is the Einstein-Hilbert Lagrangian. Because the two Lagrangians are different only by a total derivative term, which does not contribute to the equations of motion, their equations of motion are thus equivalent. This is why the theory is dubbed TEGR.

By varying the Lagrangian \eqref{Ltegr} with respect to the tetrad $e^a_\mu$, the field equation of TEGR can be obtained as
\begin{align}
    4\partial_\sigma \left( e {S_a}^{\mu\sigma} \right) - 4e {S_c}^{\nu\mu} {T^c}_{\nu a} + e e_a^\mu T - 2 \kappa^2 e \mathcal{T}_a^\mu = 0 \,,\label{classiceom}
\end{align}
where
\begin{align}
    \mathcal{T}_a^\mu = -\frac{1}{e} \frac{\delta ( e L_m )}{\delta e^a_\mu}.\label{emtt}
\end{align}
Note that $\mathcal{T}_a^\mu$ can be converted into the energy-momentum tensor expressed with tensor indices by the relation
$\mathcal{T}_\nu^\mu = e^a_\nu \mathcal{T}_a^\mu$.

\section{Effective Lagrangian with Quantum Tetrad Corrections}\label{sec.Qtegr}

As we have mentioned in the Introduction, Refs.~\cite{Dzhunushaliev:2013nea,Dzhunushaliev:2015mva} proposed an interesting approach of incorporating quantum corrections, at the semi-classical regime, into the classical gravitational theory. This approach is based on the decomposition of the metric operator into a classical part and a quantum part. By omitting higher-order contributions of the quantum part and assuming that the expectation value of the quantum part of the metric can be described by a tensor constructed solely by some classical entities, the resulting effective theory turns out to be a modified theory of gravity with non-minimal curvature-matter couplings. In this section, we will adopt this approach and extend it to the teleparallel formulation.

Following the idea of quantum decomposition of the spacetime metric proposed in Refs.~\cite{Dzhunushaliev:2013nea,Dzhunushaliev:2015mva}, in the teleparallel formulation, we assume that the tetrad field is quantum in nature, namely, it can be expressed as a quantum operator $\widehat{e_\mu^a}$. Then, we decompose the quantum tetrad field into two parts: The first part is the classical part, and the second part represents a quantum fluctuating part $\widehat{\delta e_\mu^a}$. More explicitly, we have
\begin{align}
    \widehat{e_\mu^a} = e_\mu^a + \widehat{\delta e_\mu^a} \,.
\end{align}
Then, we assume that, in general, the average of the quantum fluctuating part of the tetrad is not zero, and can be expressed as an arbitrary tetrad-like field, which is constructed solely by the classical entities in the theory:
\begin{align}
    \langle \widehat{\delta e_\mu^a} \rangle = Q_\mu^a \neq 0 \,.
\end{align}
Adopting the semi-classical approximation, i.e., neglecting higher-order contributions from the quantum tetrad, we shall replace the standard tetrad field in TEGR action with the quantum tetrad $\widehat{e_\mu^a}$ and construct the effective action encoding these quantum corrections.

In the semi-classical approximation, we consider the quantum tetrad and expand the operator version of the TEGR gravitational
Lagrangian as follows
\begin{align}
    \widehat{\mathcal{L}} &= \frac{1}{2\kappa^2} \widehat{e} \widehat{T} \approx \frac{1}{2\kappa^2} e T + \frac{1}{2\kappa^2} \frac{\delta \left( e T \right)}{\delta e^a_\rho}  \widehat{\delta e^a_\rho} \,.
    \label{3.3}
\end{align}
On the other hand, the effective matter Lagrangian is given by
\begin{align}
\label{3.4}
    \widehat{e} \widehat{\mathcal{L}}_m &\approx e \mathcal{L}_m + \frac{\delta (e \mathcal{L}_m)}{\delta e^a_\rho}  \widehat{\delta e^a_\rho}
    \,.
\end{align}
One can see that the leading order terms in the expansions \eqref{3.3} and \eqref{3.4} are given by their classical counterparts. The coefficients in the next-to-leading order terms, on the other hand, contain the variations of $eT$ and $e\mathcal{L}_m$ with respect to $e_\rho^a$, which give classical equations of motion and are shown in Eq.~\eqref{classiceom}. Therefore, the Lagrangians \eqref{3.3} and \eqref{3.4} can be expressed respectively as
\begin{align}
    &\widehat{\mathcal{L}}= \frac{1}{2\kappa^2} e T + \frac{e}{2\kappa^2} \bigg[ \frac{4}{e} \partial_\sigma \left( e {S_a}^{\rho\sigma} \right) 
    \nonumber\\
    &- 4{S_c}^{\nu\rho} {T^c}_{\nu \lambda} e_a^\lambda + e_a^\rho T \bigg]  \widehat{\delta e^a_\rho} 
\end{align}
and
\begin{align}
    \widehat{e} \widehat{\mathcal{L}}_m = e \mathcal{L}_m - e \mathcal{T}_a^\rho \widehat{\delta e^a_\rho} \,.
\end{align}

Combining the gravitational Lagrangian with matter Lagrangian and taking their expectation values, we derive the effective teleparallel Lagrangian
\begin{align}
    \langle \widehat{\mathcal{L}}+\widehat{e} \widehat{\mathcal{L}}_m \rangle= \mathcal{L}_{\text{eff}} &= \frac{1}{2\kappa^2} e T + e \mathcal{L}_m + \frac{e}{2\kappa^2} \bigg[ \frac{4}{e} \partial_\sigma \left( e {S_a}^{\rho\sigma} \right)
    \nonumber\\
    &- 4{S_c}^{\nu\rho}  {T^c}_{\nu \lambda} e_a^\lambda + e_a^\rho T - 2 \kappa^2 \mathcal{T}_a^\rho \bigg] Q^a_\rho \,.\label{qcorrection}
\end{align}
The first two terms give the classical equation of motion, while the other terms collectively encode the quantum corrections. At this point, the quantum tetrad $Q_\mu^a$ remains arbitrary. We will assume it to be constructed solely by some combinations of the classical entities. In general, there can be many possibilities. However, as we will show later, a general feature of the effective theories built with different $Q_\mu^a$ is the presence of non-minimal torsion-matter couplings. In addition, the presence of the terms associated with $Q_\mu^a$ in Eq.~\eqref{qcorrection}, i.e., the non-minimal torsion-matter couplings, would naturally lead to the non-conservation of energy-momentum tensor.

\section{Some choices for $Q_\mu^a$}\label{sec.example}

After introducing the basic setup of the effective theory, in this section, we will consider some choices of the quantum tetrad $Q_\mu^a$ and demonstrate how one can construct effective theories from these tetrads in a more explicit manner. Though the choice of the quantum tetrad $Q_\mu^a$ seems to be arbitrary, a natural requirement for choosing it is that $Q_\mu^a$ should approach zero in the absence of gravity when the torsion and energy-momentum tensor are very small i.e. negligible quantum corrections in Minkowskian spacetimes. As we will exhibit later, a common feature of these different choices is that the resulting effective theories all contain non-minimal torsion-matter couplings.

\subsection{Metric-induced corrections}
\label{subsec:scalar}
We first consider the simplest choice of the quantum tetrad $Q_\mu^a$, which is assumed to be proportional to the tetrad field $e_\mu^a$:
\begin{align}
    Q^a_\mu = \alpha \phi e^a_\mu\,,\label{choice1}
\end{align}
where $\alpha$ is a constant. Here, we allow the proportionality factor to be governed by some dynamical scalar field $\phi$. As we have mentioned, the value of the scalar field is required to be small in the absence of gravity. Similar quantum corrections were also proposed in the Riemannian formulation in Ref.~\cite{Liu:2016qfx}. A direct consequence of this choice is that the quantum tetrad directly introduces a quantum metric conformal to the classical one. More explicitly, one can construct a quantum metric using the quantum tetrad \eqref{choice1}, and its expectation value can be written as
\begin{align}
    \langle\widehat{g_{\mu\nu}}\rangle = \langle\eta_{ab} \widehat{e^a_\mu} \widehat{e^b_\nu}\rangle \approx g_{\mu\nu} ( 1 + 2\alpha\phi ) \,.
\end{align}
Using the quantum tetrad \eqref{choice1}, the effective Lagrangian \eqref{qcorrection} becomes
\begin{align}
    \mathcal{L}_{\text{eff}} = \frac{1}{2\kappa^2} e T + e \mathcal{L}_m - \frac{e\alpha\phi}{\kappa^2} \left[ \frac{2}{e} \partial_\sigma ( e {S_a}^{\sigma\rho} ) e^a_\rho + \kappa^2 \mathcal{T}  \right]\,,
\end{align}
where $\mathcal{T} \equiv \mathcal{T}^\rho_\rho$ is the trace of the energy-momentum tensor. 

To proceed, we use the following identity:
\begin{equation}
    \partial_\sigma (e {S_a}^{\sigma\rho}) e^a_\rho=\frac{e}{2} (B - T) \,,
\end{equation}
where the boundary term $B$ is defined by 
\begin{align}
    B \equiv \frac{2}{e} \partial_\sigma ( e {T_\rho}^{\rho\sigma})\,,
\end{align}
and recall that it satisfies $B - T = \mathring{R}$, where $\mathring{R}$ is the Ricci scalar defined within the standard Riemannian representation. The effective Lagrangian thus becomes
\begin{align}
\label{fTBTPHI}
    \mathcal{L}_{\text{eff}} = \frac{1}{2} e \left[ T + 2 \alpha \phi (T - B ) + 2  \mathcal{L}_m - 2 \alpha \phi \mathcal{T} \right]\,,
\end{align}
where we have set $\kappa=1$. One can see that an additional matter coupling enters the effective Lagrangian through the trace of the energy-momentum tensor. Furthermore, with the relation $B - T = \mathring{R}$, the above effective Lagrangian can be expressed in the standard Riemannian representation:
\begin{equation}
\mathcal{L}_{\text{eff}}=\frac{1}{2}\sqrt{-g}\left[-\mathring{R}-2\alpha\phi \mathring{R}+2\mathcal{L}_m-2\alpha\phi\mathcal{T}\right]\,,
\end{equation}
which is equivalent to the effective theory obtained in the Riemannian formulation by assuming that the expectation value of the quantum metric is conformal to its classical counterpart (see Eq.~(21) in Ref.~\cite{Liu:2016qfx}). This is a naive manifestation of the equivalence of GR between the Riemannian and the teleparallel formulations.

\subsection{Energy-momentum induced corrections}
The second possibility is to assume that the correction term contributed by the quantum tetrad is proportional to the energy-momentum tensor. This can be achieved by introducing the quantum tetrad $Q_\mu^a$ such that
\begin{align}
\langle\widehat{g_{\mu\nu}}\rangle&=\langle\eta_{ab}\widehat{e_\mu^a}\widehat{e_\nu^b}\rangle\nonumber\\&\approx g_{\mu\nu}+\eta_{ab}\left(e_\mu^aQ_\nu^b+e_\nu^bQ_\mu^a\right)=g_{\mu\nu}+\alpha \mathcal{T}_{\mu\nu}\,.\label{etansatz}
\end{align}
From the last equality of Eq.~\eqref{etansatz}, one gets 
\begin{equation}
\alpha\mathcal{T}_\nu^\rho=e_a^\lambda Q_\mu^a\left(\delta_\lambda^\rho \delta_\nu^\mu+g_{\lambda\nu}g^{\mu\rho}\right)\,.\label{atansatz2}
\end{equation}
If one considers an arbitrary tensor ${W_{\mu\alpha_1\alpha_2...\alpha_m}}^{\nu\beta_1\beta_2...\beta_n}$, and contracts it with Eq.~\eqref{atansatz2}, we have the identity
\begin{align}
&\alpha\mathcal{T}_\nu^\rho {W_{\rho\alpha_1\alpha_2...\alpha_m}}^{\nu\beta_1\beta_2...\beta_n}
\\ \nonumber
&=e_a^\lambda Q_\mu^a\left({W_{\lambda\alpha_1\alpha_2...\alpha_m}}^{\mu\beta_1\beta_2...\beta_n}+{{W^\mu}_{\alpha_1\alpha_2...\alpha_m\lambda}}^{\beta_1\beta_2...\beta_n}\right)\,.
\end{align}
In particular, for a symmetric tensor ${W_\mu}^\nu={W^\nu}_\mu$, this identity implies
\begin{equation}
\alpha\mathcal{T}_\nu^\rho W_\rho^\nu=2e_a^\lambda Q_\mu^aW_\lambda^\mu\,.
\end{equation}
One then gets the following useful formulae
\begin{align}
e_a^\lambda Q_\lambda^a&=\frac{1}{2}\alpha\mathcal{T}\,,\nonumber\\
e_a^\lambda Q_\mu^a\mathcal{T}_\lambda^\mu&=\frac{1}{2}\alpha\mathcal{T}_\nu^\rho\mathcal{T}_\rho^\nu\,,\label{identity12}
\end{align}
and
\begin{align}
&\left[e_a^\lambda{T^\rho}_{\mu\lambda}{S_{\rho}}^{\nu\mu}-\frac{1}{e}\partial_\mu\left(e{S_a}^{\mu\nu}\right)\right]Q_{\nu}^a=\frac{1}{4}\alpha G_{\mu\nu}\mathcal{T}^{\mu\nu}-\frac{1}{8}\alpha T\mathcal{T}\nonumber\\
=&\,\frac{\alpha}{2}\left[{T^\sigma}_{\mu\nu}{S_\sigma}^{\lambda\mu}-\frac{1}{e}e_\nu^a\partial_\mu\left(e {S_a}^{\mu\lambda}\right)\right]\mathcal{T}_\lambda^\nu\,.\label{identity3}
\end{align}
Note that on the above equation \eqref{identity3}, we have used the identity
\begin{equation}
G_\nu^\lambda=\frac{1}{2}T\delta_\nu^\lambda+2{T^\sigma}_{\mu\nu}{S_\sigma}^{\lambda\mu}-\frac{2}{e}e_\nu^a\partial_\mu\left(e{S_a}^{\mu\lambda}\right)\,.
\end{equation}

Finally, combining Eqs.~\eqref{identity12} and \eqref{identity3}, the effective Lagrangian \eqref{qcorrection} can be written as
\begin{align}
\mathcal{L}_{\text{eff}}&=\frac{1}{2}eT+e\mathcal{L}_m +\frac{\alpha e}{2}\bigg\{\frac{1}{2}T\mathcal{T}-\mathcal{T}_\mu^\nu\mathcal{T}_\nu^\mu
\nonumber\\
&+2\left[{T^\sigma}_{\mu\nu}{S_\sigma}^{\lambda\mu}-\frac{1}{e}e_\nu^a\partial_\mu\left(e {S_a}^{\mu\lambda}\right)\right]\mathcal{T}_\lambda^\nu\bigg\}\,.\label{effective5}
\end{align}
Similar to what we have shown at the end of Sec.~\ref{subsec:scalar}, using the relation $B-T=\mathring{R}$, the effective Lagrangian can also be expressed in the standard Riemannian representation:
\begin{align}
\mathcal{L}_{\text{eff}}&=\frac{1}{2}\sqrt{-g}\left[-\mathring{R}\left(1+\frac{\alpha}{2}\mathcal{T}\right)+\alpha \left(\mathring{R}_{\mu\nu}-\mathcal{T}_{\mu\nu}\right)\mathcal{T}^{\mu\nu}\right]\nonumber\\
&+\sqrt{-g}\mathcal{L}_m\,.
\end{align}
This effective Lagrangian is equivalent to the one obtained in the standard Riemannian formulation by assuming that the expectation value of the quantum metric is proportional to the energy-momentum tensor (see Eq.~(97) of Ref.~\cite{Liu:2016qfx}). This is again a manifestation of the equivalence of GR and TEGR, showing that the equivalence between the two can be extended to this semi-classical description.

\subsection{Torsion scalar-induced corrections}
In the previous two subsections, we have constructed two effective Lagrangians in the teleparallel formulation by assuming two different quantum tetrads $Q_\mu^a$. One common feature is the appearance of non-minimal torsion-matter couplings in the effective Lagrangian. We have also shown that the effective Lagrangians are equivalent to their standard Riemannian counterparts, indicating that the equivalence between GR and TEGR can be extended to the semi-classical description proposed in Ref.~\cite{Dzhunushaliev:2013nea} and this paper. A natural question then arises: Can one construct a novel effective Lagrangian in the teleparallel formulation? In this subsection, we will show explicitly that this is indeed possible.

We shall choose the following quantum tetrad
\begin{align}
    Q^a_\mu = \alpha T e^a_\mu\,,
\end{align}
where we have simply replaced the dynamical scalar field $\phi$ in the effective Lagrangian of Sec.~\ref{subsec:scalar} by the torsion scalar $T$. The resultant effective Lagrangian reads
\begin{align}
    \mathcal{L}_{\text{eff}} = \frac{1}{2} e \big[ T + 2 \alpha T (T - B ) + 2  \mathcal{L}_m - 2 \alpha T \mathcal{T} \big] \,.\label{f(T,B,T)}
\end{align}
This is a novel effective Lagrangian in the sense that there is no equivalent counterpart of the theory in the Riemannian formulation. In fact, the effective Lagrangian \eqref{f(T,B,T)} is a subclass of the $f(T,B,\mathcal{T})$ theory that in general contains the torsion scalar $T$, the boundary term $B$, and the non-minimal matter couplings encoded in the trace of the energy-momentum tensor $\mathcal{T}$. In Sec.~\ref{sec:cosmology}, we will investigate some interesting cosmological solutions in the theory given by the effective Lagrangian \eqref{f(T,B,T)}.

Before closing this section, we will write down the general equation of motion of the $f(T,B,\mathcal{T})$ gravity. We consider the action 
\begin{equation}
\mathcal{S}=\frac{1}{2}\int d^4x ef(T,B,\mathcal{T})+\mathcal{S}_m\,.
\end{equation}
The variation of the action gives{\footnote{For the detailed derivation of the variation of the boundary term $B$ with respect to the tetrad, we refer the readers to Ref.~\cite{Bahamonde:2015zma}.  }}:
\begin{align}
\delta\mathcal{S}&=\int d^4x\frac{1}{2}\left[ef_T\delta T+ef_B\delta B+ef_\mathcal{T}\delta\mathcal{T}+f\delta e\right]-e\mathcal{T}_a^\nu\delta e_\nu^a\nonumber\\
&=\int d^4x e\Big[\frac{1}{2}fe_a^\nu+2f_T{T^\rho}_{\mu a}{S_\rho}^{\nu\mu}-\frac{2}{e}\partial_\mu\left(ef_T{S_{a}}^{\mu\nu}\right)\nonumber\\
&+e_a^\rho\nabla^\nu\nabla_\rho f_B-e_a^\nu\Box f_B-\frac{1}{2}Bf_Be_a^\nu-2\left(\partial_\mu f_B\right){S_{a}}^{\mu\nu}\nonumber\\
&+\frac{f_\mathcal{T}}{2}\left(g^{\alpha\beta}\frac{\partial\mathcal{T}_{\alpha\beta}}{\partial e_\nu^a}-2\mathcal{T}_a^\nu\right)
 -\mathcal{T}_{a}^\nu\Big]\delta e_\nu^a\,.
\end{align}
We get the field equation
\begin{align}
&\frac{1}{2}fe_a^\nu+2f_T{T^\rho}_{\mu a}{S_\rho}^{\nu\mu}-\frac{2}{e}\partial_\mu\left(ef_T{S_{a}}^{\mu\nu}\right)+e_a^\rho\nabla^\nu\nabla_\rho f_B\nonumber\\
-&\,e_a^\nu\Box f_B-\frac{1}{2}Bf_Be_a^\nu-2\left(\partial_\mu f_B\right){S_{a}}^{\mu\nu}\nonumber\\=&-\frac{f_\mathcal{T}}{2}\left(g^{\alpha\beta}\frac{\partial\mathcal{T}_{\alpha\beta}}{\partial e_\nu^a}-2\mathcal{T}_a^\nu\right)+\mathcal{T}_{a}^\nu\,.
\end{align}
One can see that when the coupling of the boundary term $B$ is turned off, the theory reduces to the $f(T,\mathcal{T})$ theory, which was firstly proposed in Ref.~\cite{Harko:2014aja}{\footnote{In order to be consistent with the notation used in \cite{Harko:2014aja}, throughout this paper, we have used $\mathcal{T}^\mu_\nu$ and $\mathcal{T}$ to represent the energy-momentum tensor and its trace, instead of $\Theta^\mu_\nu$ and $\Theta$.}. Furthermore, if the functional form of $f$ can be written as $f(T,B,\mathcal{T})=f(T-B,\mathcal{T})$, the theory can be recast into the Riemannian formulation, and the $f(\mathring{R},\mathcal{T})$ gravity is recovered.

It should be emphasized that very recently, a novel theory which contains non-minimal torsion-boundary-matter couplings has been proposed in Ref.~\cite{Bahamonde:2017ifa}, which is called $f(T,B,\mathcal{L}_m)$ gravity. The $f(T,B,\mathcal{T})$ theory is in general different from the $f(T,B,\mathcal{L}_m)$ theory due to the different forms of non-minimal matter couplings.

We would like to also mention that the quantum corrections and the associated non-minimal geometry-matter couplings discussed in this paper can have various cosmological applications, such as inflation, primordial bouncing universe, and late-time universe. Black hole spacetimes could also receive these quantum corrections. In Riemannian formulation, these applications have been discussed in Refs.~\cite{Yang:2015jla,Liu:2016qfx,Yang:2020lxv}. We expect that similar effects would also appear in the teleparallel formulation, in particular for the models that have Riemannian counterparts discussed in Refs.~\cite{Yang:2015jla,Liu:2016qfx,Yang:2020lxv}. As for the novel model \ref{f(T,B,T)}, the non-minimal torsion-boundary-matter couplings would naturally lead to more non-trivial effects, in particular, in strong gravity regimes.

\section{Cosmological solutions}
\label{sec:cosmology}

In this section, we will focus on the novel effective Lagrangian \eqref{f(T,B,T)} and investigate its cosmological solutions. In particular, we will assume that the matter field is governed by a dynamical scalar field $\phi$, whose matter Lagrangian is given by
\begin{equation}
\mathcal{L}_m=\frac{1}{2}\partial_\mu\phi\partial^\mu\phi-V(\phi)\,,
\end{equation}
where $V(\phi)$ is the scalar field potential. In cosmology, the scalar field model can be used to describe the early inflationary stage (inflaton), or the late-time accelerating expansion (dark energy).

In the following, we will consider the effective Lagrangian \eqref{f(T,B,T)}, and assume that the universe is flat, homogeneous and isotropic on its largest scale. The metric is described by the flat Friedmann-Robertson-Walker (FRW) metric
\begin{align}
    ds^2 = N^2(t) dt^2 - a^2(t) \delta_{ij} dx^i dx^j\,,
\end{align}
where $N(t)$ is the lapse function and $a(t)$ is the scale factor, which are functions of the cosmic time $t$. The corresponding tetrad can be written as
\begin{align}
\label{tetrad}
    e^a_\mu = \text{diag} (N(t),a(t),a(t),a(t))\,,
\end{align}
where the tetrad is compatible with the Weitzenb\"ock gauge which allows us to take a vanishing spin connection. To derive the cosmological equations, we are going to construct the minisuperspace model for the cosmological system. Under the choice of the tetrad \eqref{tetrad}, the torsion scalar $T$ and the boundary term $B$ can be expressed as:
\begin{align}
    T = - 6H^2\,, \quad B = - 6 \left[ 3H^2 + \frac{\dot{H}}{N} \right]\,,
\end{align}
respectively, where $H = H(t) \equiv \dot{a}/(a N)$ is the Hubble function. The dot denotes the derivative with respect to the cosmic time $t$. Furthermore, the trace of the energy-momentum tensor of the scalar field $\phi(t)$ can be written as
\begin{align}
    \mathcal{T} = - \frac{\dot{\phi}^2}{N^2} + 4 V\,.
\end{align}
The effective Lagrangian of this minisuperspace model \eqref{f(T,B,T)} can then be expressed as
\begin{align}
    &\mathcal{L}_{\text{eff}} = \frac{1}{2 a N^4} \bigg[ -72 \alpha  \dot{a}^4 N + 72 \alpha a \dot{a}^2 ( \dot{a} \dot{N} - \ddot{a} N ) 
    \nonumber\\ 
    &+ a^4 (  N^3 \dot{\phi}^2  - 2 N^5 V ) + 6 a^2 \dot{a}^2 N ( N^2( 8 \alpha V - 1) - 2 \alpha \dot{\phi}^2 )  \bigg] \,.\label{minil}
\end{align}
The equations of motion can be obtained by varying the Lagrangian \eqref{minil}. By varying the effective Lagrangian with respect to $N$, $a$, and $\phi$, one can obtain the effective Friedmann equation, Raychaudhuri equation, and the modified Klein-Gordon equation, respectively. By assuming $N = 1$, the effective Friedmann equation and the Raychaudhuri equation can be written as
\begin{align}
    3 H^2 &= \rho_{\text{eff}}\,,\label{effectiveFriedmann}\\
    3 H^2 + 2 \dot{H} &= - p_{\text{eff}}\label{effectiveRay}\,,
\end{align}
respectively, where the effective energy density and pressure are defined as
\begin{align}
    \rho_{\text{eff}} &= \frac{1}{2} \dot{\phi}^2 ( 1 - 36 H^2 \alpha) + ( 1 + 24 H^2 \alpha) V - 108 \alpha H^4\,,\\
    -p_{\text{eff}} &= - \frac{1}{2} \dot{\phi}^2 ( 1 + 12 H^2 \alpha + 8 \alpha \dot{H}) 
   \nonumber \\ 
    &+ (1 + 24 H^2 \alpha + 16 \alpha \dot{H}) V + 8 \alpha H \dot{\phi} (2 V_{\phi} - \ddot{\phi}) 
    \nonumber\\ 
    &- 144 \alpha H^2 \dot{H} - 108 \alpha H^4 \,.
\end{align}
The effective energy density and pressure contain both the matter contribution from the scalar field, as well as the geometric contribution from the torsion and the boundary term. One can further define the effective equation of state as follows
\begin{equation}
    w_{\text{eff}} = \frac{p_{\text{eff}}}{\rho_{\text{eff}}} \,.\label{defeff}
\end{equation}
Finally, the modified Klein-Gordon equation is
\begin{align}
\label{effectiveKG}
    3 H \dot{\phi} ( 1 &- 12 \alpha H^2 - 8 \alpha \dot{H} )
     \nonumber\\
    &+ V_{\phi} ( 1 - 24 \alpha H^2) + \ddot{\phi} ( 1 - 12 \alpha H^2 ) = 0 \,,
\end{align}
where $V_\phi\equiv dV/d\phi$. Essentially, the Klein-Gordon equation is modified compared with that in GR because of the existence of the non-minimal matter couplings in the theory. A direct consequence of these couplings is that the energy-momentum tensor is not covariantly conserved, giving rise to energy transfers between gravity (torsion and boundary terms) and the cosmological fluids (scalar field) in the universe.

\subsection{Constant Hubble function}

We first start with the assumption that the Hubble function is a positive constant, namely, a pure de Sitter universe, to see wether the theory allows such an accelerating cosmic expansion. We consider the ansatz
\begin{equation}
a(t)=a_0e^{H_0t}\,,\qquad H(t)=H_0\,.
\end{equation}
Plugging this ansatz into the equations of motion, we obtain the following equations
\begin{align}
\dot\phi&=0\,,\qquad V_\phi=0\,,\\
V&=\frac{3H_0^2\left(1+36\alpha H_0^2\right)}{1+24\alpha H_0^2}\,.\label{constantHV}
\end{align}
One can see that the effective theory \eqref{f(T,B,T)} with a dynamical scalar field is able to generate a de Sitter universe when $\phi$ approaches a constant at a local extrema of the potential. Essentially, given a constant Hubble function $H_0$, the value of the corresponding potential on which the scalar field resides can be uniquely determined using Eq.~\eqref{constantHV}. 

Some interesting results for a negative $\alpha$ have to be pointed out. First, when $|\alpha|H_0^2=1/36$, the universe is essentially de Sitter, but the potential value is zero. This is solely due to the existence of the quartic term of the Hubble function, i.e., $H^4$, in the equations of motion. This higher order term is contributed by the boundary term $B$ in the theory and is induced by quantum effects, i.e., a non-zero $\alpha$.

Second, if $|\alpha|H_0^2\approx1/24$, the potential could acquire a large value. By tuning the value of $\alpha$, this scenario could happen even for a relatively small $H_0^2$. This property could shed some light on the resolution of the cosmological constant problem. The detailed investigation in light of this direction is beyond the scope of this paper, and we will leave it for the future publications.

\subsection{Hybrid Expansion Law}

Now, we shall consider a more realistic cosmic evolution and see whether such a cosmological solution can be obtained in the effective theory \eqref{f(T,B,T)}. In particular, we apply the so-called Hybrid Expansion Law (HEL) \cite{Akarsu:2013xha} for the cosmic evolution
\begin{align}
\label{HEL}
    a(t) = a_0 \left( \frac{t}{t_0} \right)^\gamma e^{\beta \left( \frac{t}{t_0} - 1 \right)}\,,
\end{align}
where $\gamma$ and $\beta$ are free parameters. On the above ansatz, $a_0$ and $t_0$ are the scale factor and the cosmic age at the present time, respectively. The corresponding Hubble function is
\begin{equation}
    H = \frac{\gamma}{t} + \frac{\beta}{t_0} \,.\label{Hpara}
\end{equation}
The HEL is a useful ansatz for the cosmic evolution because it naturally describes a decelerating power-law expansion of the universe, followed by an accelerating expansion which approaches to a de Sitter universe in the asymptotic future. The parameter $\gamma$ determines the power-law expansion at early time. On the other hand, the parameter $\beta$ controls the asymptotic value of the Hubble function. From now on, we will assume $\gamma=2/3$, such that the HEL can describe the transition from a matter-dominated universe to an asymptotic de Sitter spacetime.{\footnote{For completeness, one can include, beside the dynamical scalar field, a perfect fluid with equation of state $w=0$ to support the power-law expansion. In this work, we will focus more on the era after the transition, in which, from a phenomenological point of view, the dynamical scalar field $\phi$ is more important.}}

In FIG.~\ref{H}, we show the time evolution of the Hubble function \eqref{Hpara} for different values of $\beta$. The red, blue, black, and orange curves show the evolution for $\beta=1$, $1.2$, $1.5$, and $1.8$, respectively. Furthermore, in FIG.~\ref{weff}, we show the time evolution of the effective equation of state defined in Eq.~\eqref{defeff}. For the HEL ansatz, the effective equation of state $w_{\text{eff}}$ can be expressed explicitly as
\begin{equation}
w_{\text{eff}} = \frac{2\gamma}{3t^2} \left( \frac{\gamma}{t} + \frac{\beta}{t_0} \right)^{-2} - 1 \,.\label{eff518}
\end{equation}
Notice that in the HEL ansatz, the Hubble function and the effective equation of state are completely determined by $\gamma$ and $\beta$. From FIG.~\ref{weff}, one can see that when $t\ll t_0$, the effective equation of state approaches zero, indicating a matter dominated universe. On the other hand, when $t\gg t_0$, the universe approaches a de Sitter expansion with $w_{\text{eff}}\rightarrow-1$.

After inserting the HEL ansatz, we will reconstruct the scalar field and its potential by solving Eqs.~\eqref{effectiveRay} and \eqref{effectiveKG} numerically. The modified Friedmann equation \eqref{effectiveFriedmann} is a constraint equation and we use it to confirm the validity of our numerical solutions. In order to solve the equations, we have to specify the values of $\alpha$. The evolution of the scalar field $\phi(t)$ and the associated potential profile $V(\phi)$ are shown in FIG.~\ref{Plotphi} and FIG.~\ref{Vt}, respectively, with $\alpha=0.1$ (solid) and $\alpha=1$ (dashed). 

\begin{figure}[!ht]
\centering
\vspace{-3pt}
\includegraphics[width = .4\textwidth]{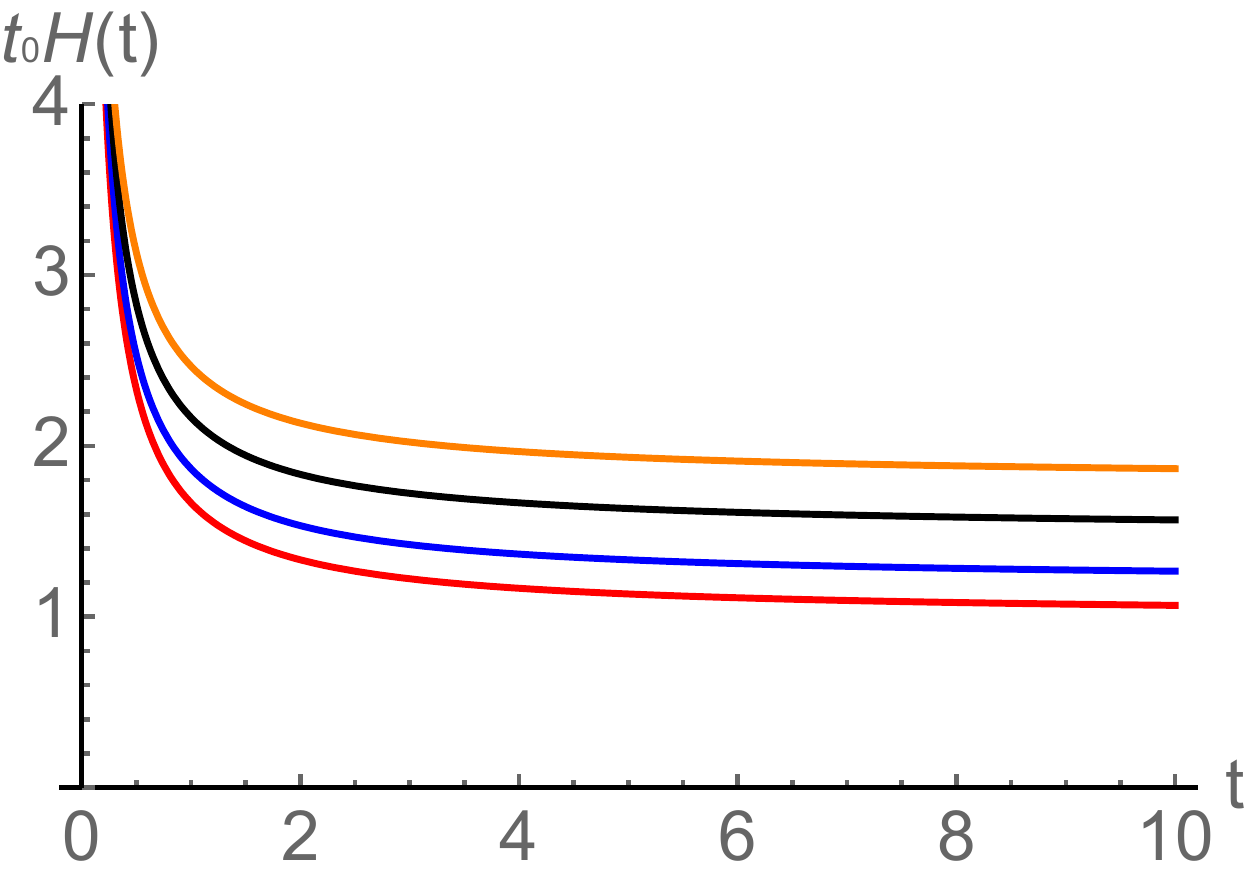}
\vspace{-10pt}
\caption{\label{H}The Hubble function $H(t)$ given by Eq.~\eqref{Hpara} as a function of $t/t_0$. The red, blue, black, and orange curves show the Hubble function for $\beta=1$, $1.2$, $1.5$, and $1.8$, respectively. In this figure, we assume $\gamma=2/3$.}
\end{figure}

\begin{figure}[!ht]
\centering
\vspace{-3pt}
\includegraphics[width = .4\textwidth]{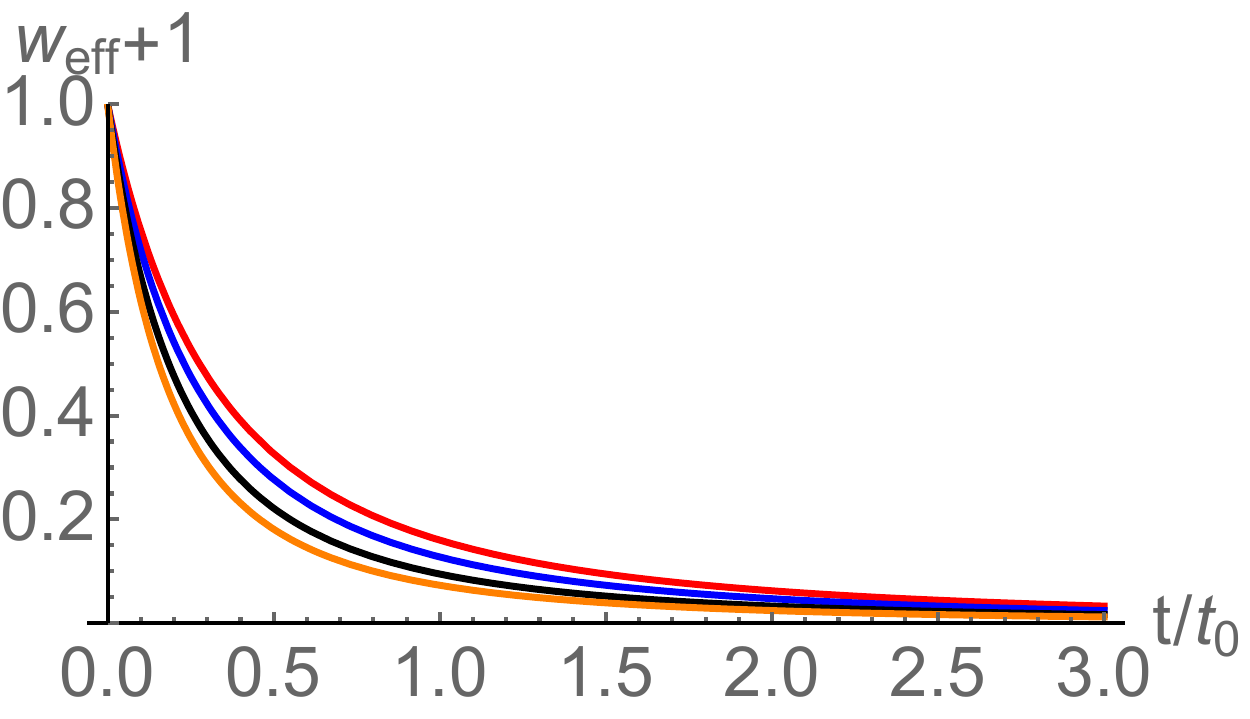}
\vspace{-10pt}
\caption{\label{weff}The effective equation of state $w_\textrm{eff}$ given by Eq.~\eqref{eff518} as a function of $t/t_0$. The parameters corresponding to each curve are the same as those in FIG.~\ref{H}.}
\end{figure}

According to FIGs.~\ref{Plotphi} and \ref{Vt}, the scalar field and its potential in this effective model can describe the smooth transition from a decelerating expansion to a late-time accelerating expansion. Different values of $\beta$ in the HEL ansatz correspond to different asymptotic expansion rates. Essentially, the scalar field approaches a constant value when $t\rightarrow\infty$, namely, its kinetic energy vanishes asymptotically. In addition, when the scalar field approaches its asymptotic value, the slope of the potential, i.e., $V_\phi$, gradually vanishes. Of course, different choices of $\alpha$ would give distinctive quantitative asymptotic values of the field. However, the qualitative behaviors of the field evolution are basically the same. We observe that the effects of changing $\alpha$ are smaller if the parameter $\beta$ is larger.

\begin{figure}[!ht]
\centering
\vspace{-3pt}
\includegraphics[width = .4\textwidth]{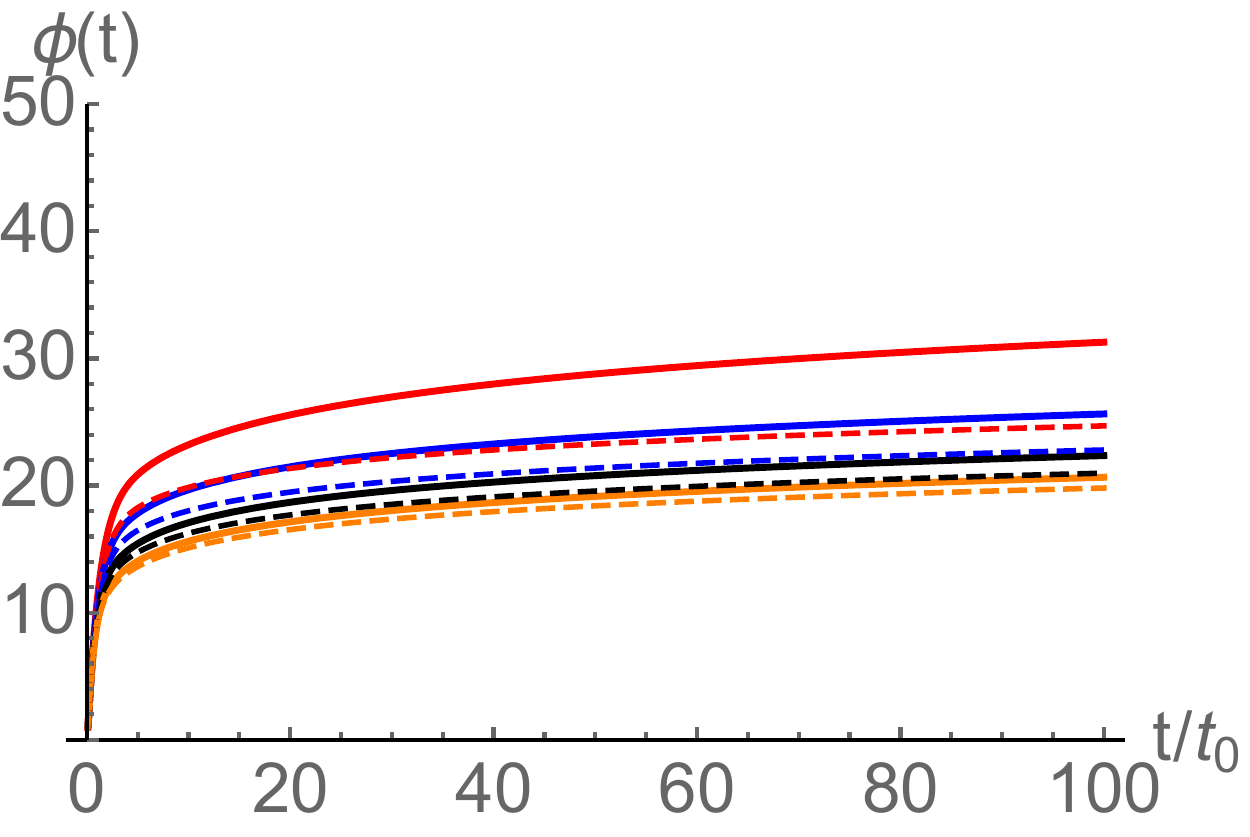}
\vspace{-10pt}
\caption{\label{Plotphi}The reconstruction of the scalar field as a function of $t/t_0$ based on the HEL \eqref{Hpara}. The solid and the dashed curves represent $\alpha=0.1$ and $\alpha=1$, respectively. The values of $\beta$ corresponding to each color are the same as those in FIG.~\ref{H}.}
\end{figure}

\begin{figure}[!ht]
\centering
\vspace{-3pt}
\includegraphics[width = .4\textwidth]{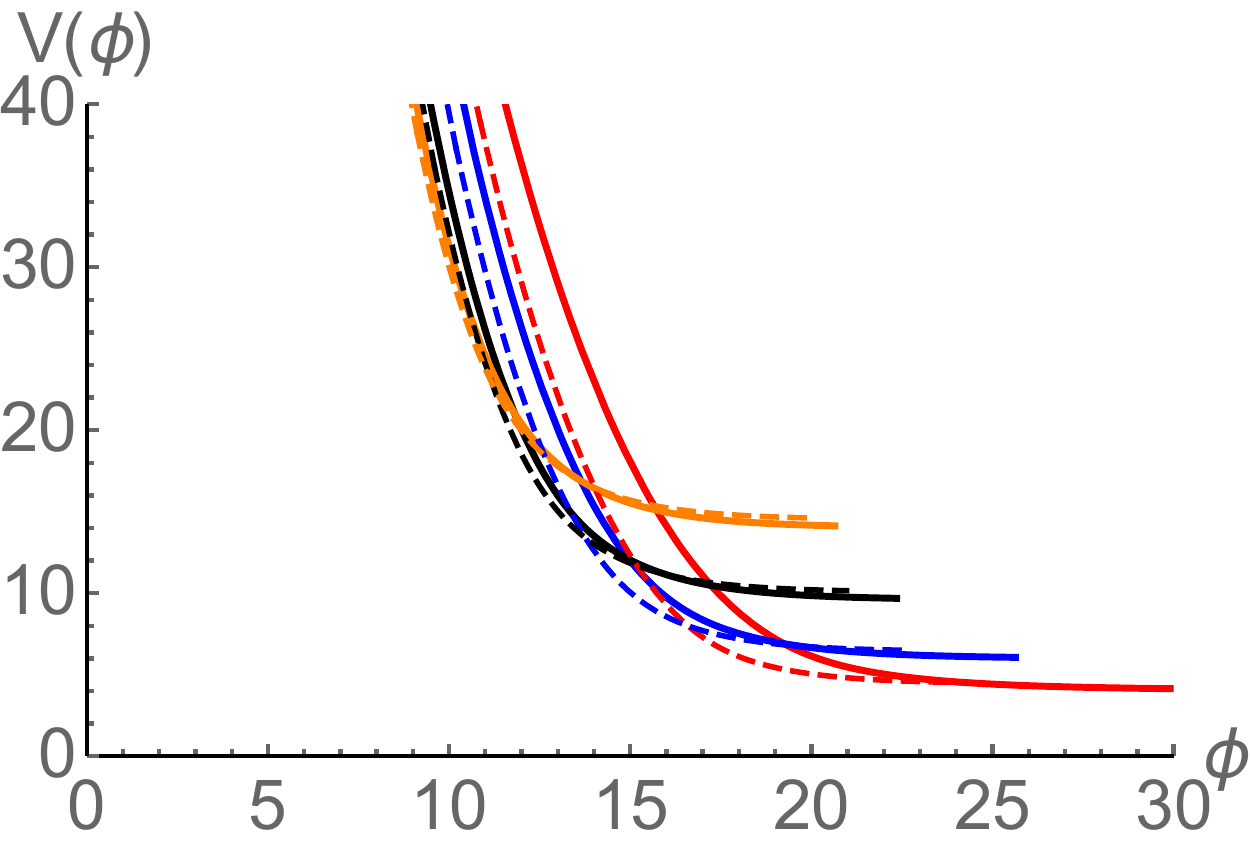}
\vspace{-10pt}
\caption{\label{Vt}The reconstruction of the scalar field potential $V(\phi)$ based on the HEL \eqref{Hpara}. The solid and the dashed curves represent $\alpha=0.1$ and $\alpha=1$, respectively. The values of $\beta$ corresponding to each color are the same as those in FIG.~\ref{H}. Each curve terminates on the right when $\phi$ approaches the value corresponding to the asymptotic de Sitter expansion.}
\end{figure}

\section{Conclusions}\label{sec.con}

Although so far a consistent formulation of a complete quantum theory of gravity is still lacking, one consensus in theoretical physics is that, no matter how quantum gravity is supposed to be formulated, the standard GR is expected to be modified in some way within quantum regimes. To evade the difficulty of complete quantization of gravity, one can incorporate quantum effects to gravity by considering semi-classical approximations. In Refs.~\cite{Dzhunushaliev:2013nea,Dzhunushaliev:2015mva}, the authors have considered the semi-classical approximations and constructed effective models of gravity by assuming that the metric operator can be decomposed into a classical and a quantum parts. Assuming that the expectation value of the quantum part can be described by some combinations of classical entities, one can obtain the corresponding semi-classical effective theories. It turns out that the resulting effective theories can be interpreted as modified theories of gravity with non-minimal curvature-matter couplings. These non-minimal couplings lead to the non-conservation of the energy-momentum tensor and they imply the existence of irreversible particle creation processes through gravity. These particle creations naturally appear in quantum field theories in curved spacetimes and there could be a more fundamental relation between quantum gravitational fields and non-minimal geometry-matter couplings.

In this paper, we address the question whether the emergence of non-minimal geometry-matter couplings is generic or not. We extend the idea of Refs.~\cite{Dzhunushaliev:2013nea,Dzhunushaliev:2015mva} to the teleparallel formulation. Analogously, we decompose the tetrad field into a classical and quantum part, then assume that the expectation value of the quantum part can be described by some classical entities. As a consequence, the resulting effective theories are all featured by the existence of non-minimal torsion-matter couplings. In addition, we have verified that, for some specific choices of the quantum tetrads, the resulting effective theories have their equivalent counterparts in the standard Riemannian formulation. In this sense, the equivalence between GR and TEGR can be extended in these semi-classical descriptions. Besides, we further consider the possibility when the quantum correction is proportional to the torsion scalar $T$ in the teleparallel gravity. This then leads to an effective theory \eqref{f(T,B,T)}, which is a subclass of the novel $f(T,B,\mathcal{T})$ theory. The novelty of this effective theory can be understood by the fact that there is no equivalent counterpart of the theory in the Riemannian formulation.

With the effective Lagrangian \eqref{f(T,B,T)}, we investigate its cosmological solutions by assuming the matter Lagrangian is given by a dynamical scalar field. In this work, we focus on the minisuperspace model of the flat FRW metric and rewrite the effective Lagrangian in terms of the scale factor, the lapse function, and the dynamical scalar field. The effects of higher-order terms of the Hubble function and the non-minimal coupling between the torsion and the scalar field naturally appear in the equations of motion. Then, we solve the Raychaudhuri equation and the Klein-Gordon equation to obtain the cosmological evolution. In this article, we first consider the ansatz of a pure de Sitter universe with a constant Hubble function $H_0$. We find that the de Sitter universe can be generated when the dynamical scalar field reaches a constant at the local extrema of the potential. The potential value at the extrema is uniquely determined by $H_0$ according to Eq.~\eqref{constantHV}. The potential depends on a free parameter $\alpha$ and can be tuned to be either zero or a large value. The latter case might moderate the cosmological constant problem.

Furthermore, we investigate the cosmic evolution in the effective theory \eqref{f(T,B,T)} by applying the ansatz of Hybrid Expansion Law (HEL). We numerically solve the evolution of the Hubble function and the dynamical scalar field such that the universe evolves smoothly from a matter-dominated phase to an asymptotic de Sitter phase. On the other hand, the numerical solution for the dynamical scalar field is found, and the scalar field approaches an asymptotic value determined by the expansion rate of the universe when $t \rightarrow \infty$. At the same time, the slope of the scalar field potential vanishes when the associated scalar field reaches its asymptotic value. We confirm the existence of the cosmological evolutions that describe a smooth transition from a decelerating expansion to an asymptotic de Sitter expansion.

In fact, non-minimal torsion-matter couplings can provide interesting phenomenology for late-time universe. In Ref.~\cite{Geng:2011aj}, it has been shown that non-minimal couplings between torsion and a scalar field could generate a phantom-like dark energy equation of state, even though the scalar field itself is non-phantom. Following this insight, we may investigate the late-time cosmology of our effective quantum model and the possible relation between dark energy and quantum fluctuations in the future. 

So far, we have only considered the models constructed by assuming that the expectation value of the quantum tetrad is non-zero. A more reasonable treatment would be to assume a zero expectation value of the quantum tetrad in its first-order, and consider the quantum corrections contributed by the product of quantum tetrads at two points. This scenario has been considered in the Riemannian formulation in Ref.~\cite{Dzhunushaliev:2015mva}. It is interesting to extend this scenario to the teleparallel formulation. In addition, in order to check the robustness of the emergence of non-minimal geometry-matter couplings when quantum corrections are included, it is necessary to extend this semi-classical approach to the Palatini formulation in which the metric and the affine connection are independent. One may also include the non-metricity \cite{BeltranJimenez:2017tkd,Harko:2018gxr,Xu:2019sbp} to see whether the similar conclusion still applies. We leave these projects to future works.

\acknowledgements
CYC is grateful to Reginald Christian Bernardo for comments on an earlier version of this draft. CYC is supported by the Institute of Physics of
Academia Sinica. YHK is supported by the National Center for Theoretical Science, Physics Division.

\end{document}